\documentclass[aps,pre,showpacs,review]{revtex4}
\usepackage[utf8]{inputenc}
\usepackage{graphicx,color}
\usepackage{subfig}
\usepackage{xcolor}
\usepackage{mathtools} 

\definecolor{verde}{rgb}{0.12, 0.3, 0.17}

\newcommand{\m}{$m^*$}
\newcommand{\M}{$M^*$}

\newcommand{\ffi}{$\phi$}
\newcommand{\etal}{\emph{et al.} }
\newcommand{\dens}{\psi (\mathbf x,\theta,t)}
\begin{document}

\title{Emergence of a single cluster in Vicsek's model at very low noise.}
\author{Lucas Barberis}
\affiliation{Instituto de Física Enrique Gaviola - CONICET, Córdoba, Argentina}
\affiliation{Universit{\'e} Nice Sophia Antipolis, Laboratoire J.A. Dieudonn{\'e}, Nice, France}
\email{lbarberis@unc.edu.ar}

\begin{abstract}
The classic Vicsek model [Phys.Rev.Lett. {\bf75},1226(1995)] is studied in the regime of very low noise intensities, which is shown to be characterized by a cluster (MC) that contains a macroscopic fraction of the system particles. It is shown that the well-known power-law behavior of the cluster size distribution loses its cutoff becoming bimodal at very low noise intensities:  A peak develops for larger sizes to settle the emergence of the MC. The average cluster number \m\, is introduced as a parameter that properly describes this change, i.e. a line in the noise-speed phase portrait can be identified to separates both regimes. The average largest cluster parameter also develops large fluctuations at a non zero critical noise. Finite size scaling analysis is performed to show that a phase transition to a macroscopic cluster is taking place. Consistency of the results with the literature is also checked and commented upon.
\end{abstract}


\maketitle

\section{Introduction}\label{s:intro}

Since its presentation in the middle 90's, Vicsek's model \cite{Vicsek1995} has become a very popular description of a system of\emph{ self-propelled particles} (SPP) and has been the workhorse of most theoretical research in this field. Its success results from being considered the minimal model able to exhibit an \emph{order-disorder transition} (ODT) and it has been implemented to study lots of biological, medical and ecological SPP problems as it becomes clear, for example, in the Visek and Zafeiris review.  \cite{Vicsek2012}

Over the last twenty years attention was focused on this ODT, thus allowing the occurrence of other interesting phenomena to go almost unnoticed. This could have happened because the ODT has been conventionally described by a \emph{ferromagnetic order parameter} that is not sensitive to the structure of the patterns developed by the model. This order parameter, is defined as: 
   
\begin{equation}\label{eq:phi}
 \phi=\frac{1}{N}\sum_{j=1}^N \exp \left( i\theta_j \right),
\end{equation}

\noindent with $N$ the number of self-propelled particles present in the system (its \emph{size}) and $\theta_j$ the orientation of the $j^{th}$ particle's velocity. The use of this order parameter is widely justified because the model implements a ferromagnetic alignment of the particle's velocity inside a short-range fixed distance, the original set of equations is:

 \begin{subequations}
 \label{eq:langevin}
 \begin{eqnarray}
 {\mathbf x}_i(t+1)&=&{\mathbf x}_i(t) + v_0 {\mathbf V}(\theta_i(t)) \times \Delta t,\label{eq:langevin1}\\
{\theta_i(t+1)}&=&\langle\theta\rangle_{i,{R_0}}+\eta \xi,\label{eq:langevin2} 
\end{eqnarray}
\end{subequations}
\noindent  where $i\in [1,N]$, and  $\langle\theta\rangle_{i,R_0}$ describes the average orientation of the neighboring particles  inside a circular region centered in the position of the $i^{th}$ particle with radius $R_0$, the interaction radius. $\Delta t$ is a discrete time step, ${\mathbf V}(\theta_i(t))$ is an unit vector pointing in the $\theta_i(t)$ direction, $\xi$ is a delta-correlated white noise of unit magnitude, and $\eta$ is the intensity of this noise, which will be a control parameter.
 
This alignment interaction is commonly called \emph{metric} and has a \emph{topological} counterpart, which restricts the alignment to neighbors defined by a Voronoi tesselation \cite{Ginelli2010} or to be fixed in number \cite{Barberis2014}, instead of being determined by an interaction radius.

\begin{figure}[h]
 \centering
  \subfloat[$\eta=0.0001$\label{fig:confMC}]{
    \includegraphics[width=0.48\textwidth]{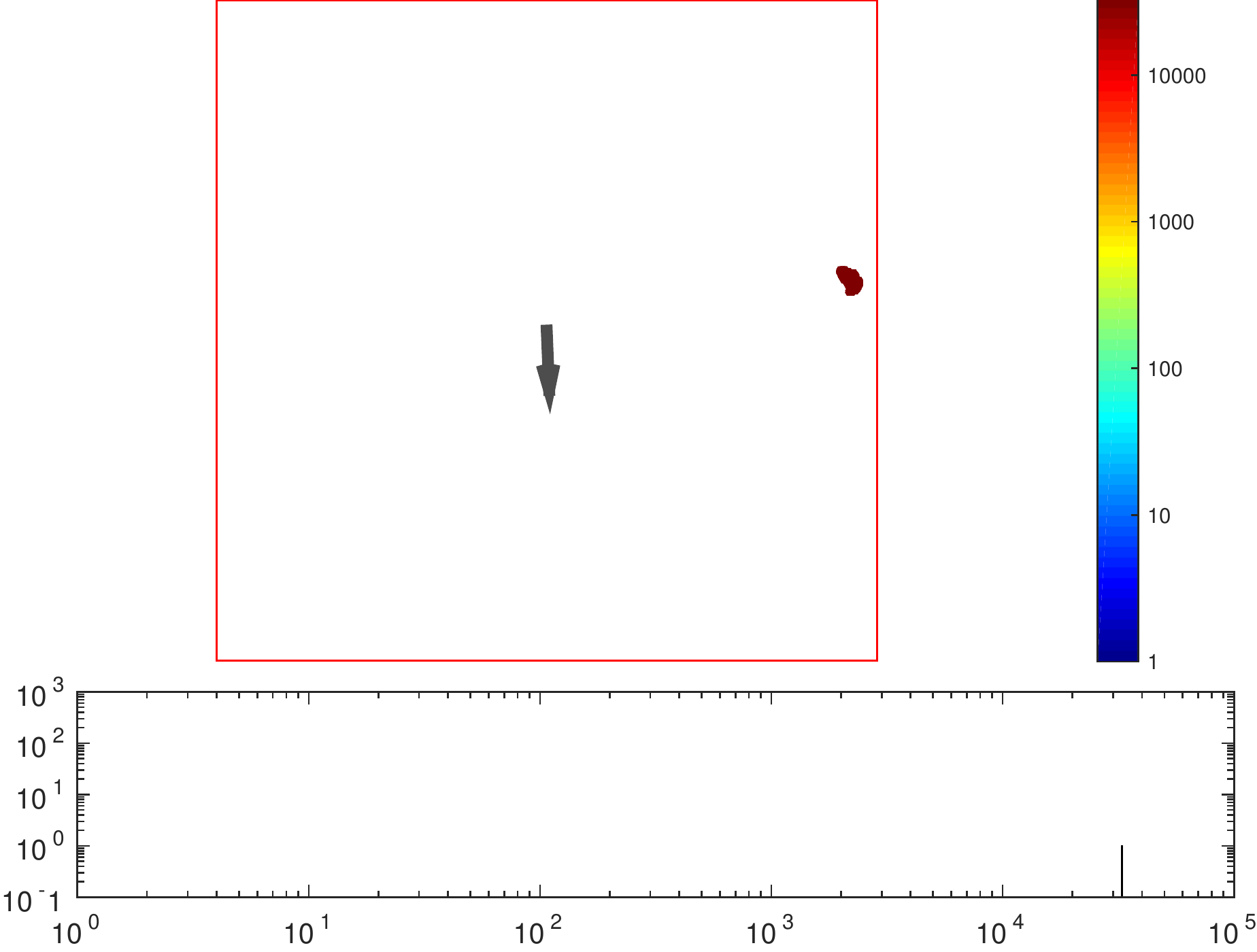} 
    }
    \hfill
   \subfloat[$\eta=0.001$\label{fig:confTF}]{
   \includegraphics[width=0.48\textwidth]{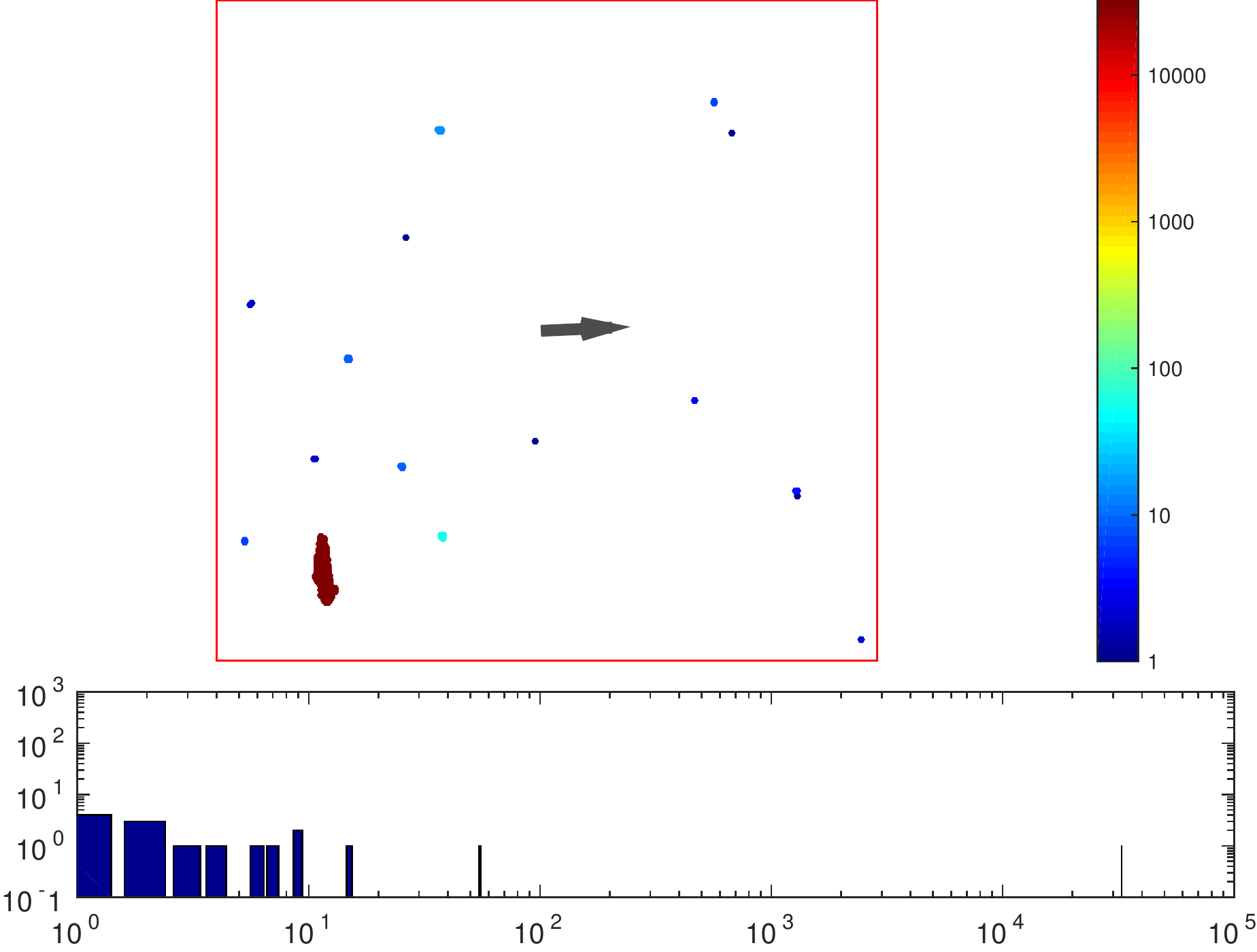}
   }
   \\
   \subfloat[$\eta=0.002$\label{fig:confTFH}]{
   \includegraphics[width=0.48\textwidth]{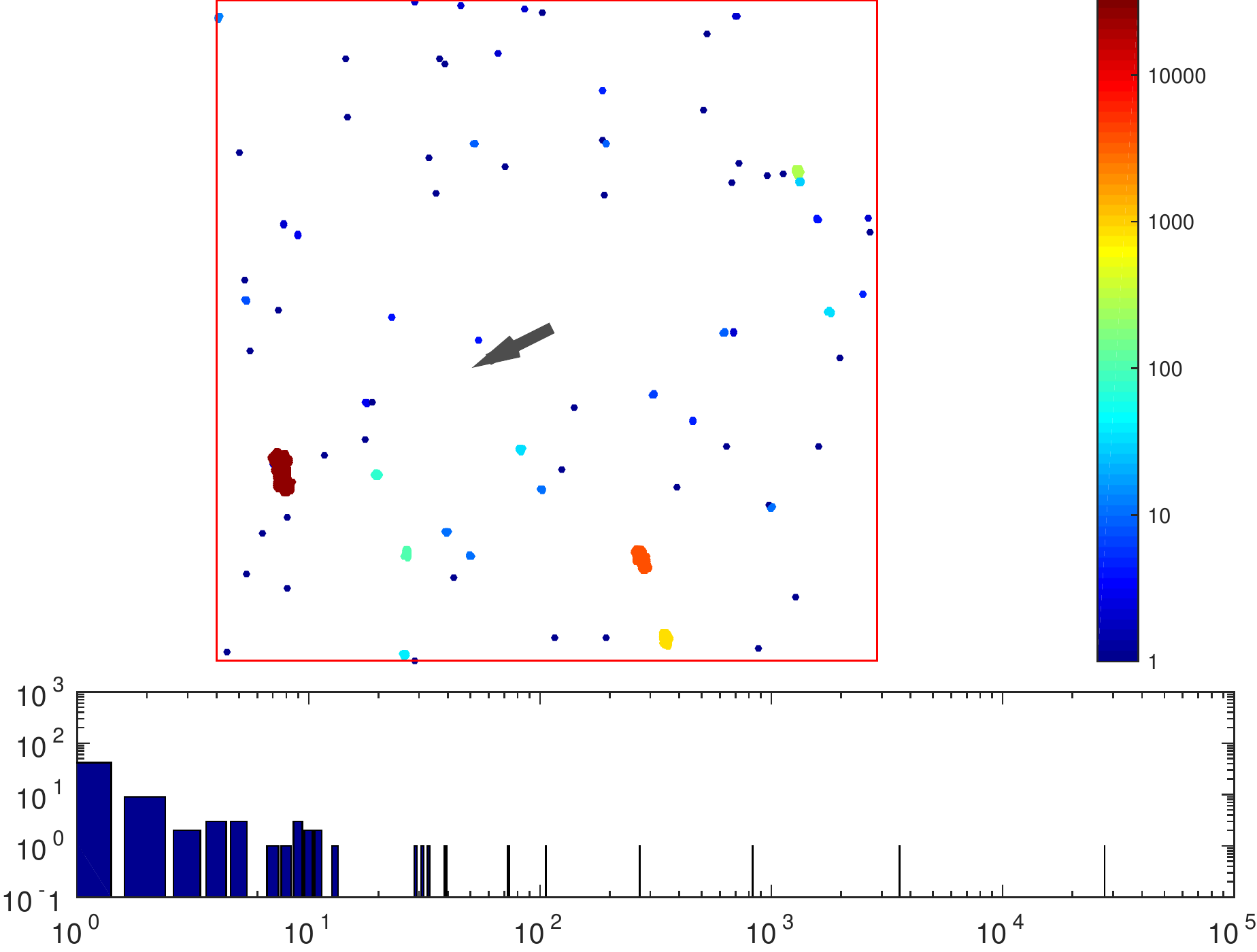}
   }
   \hfill
   \subfloat[$\eta=0.01$\label{fig:confHomo}]{
   \includegraphics[width=0.48\textwidth]{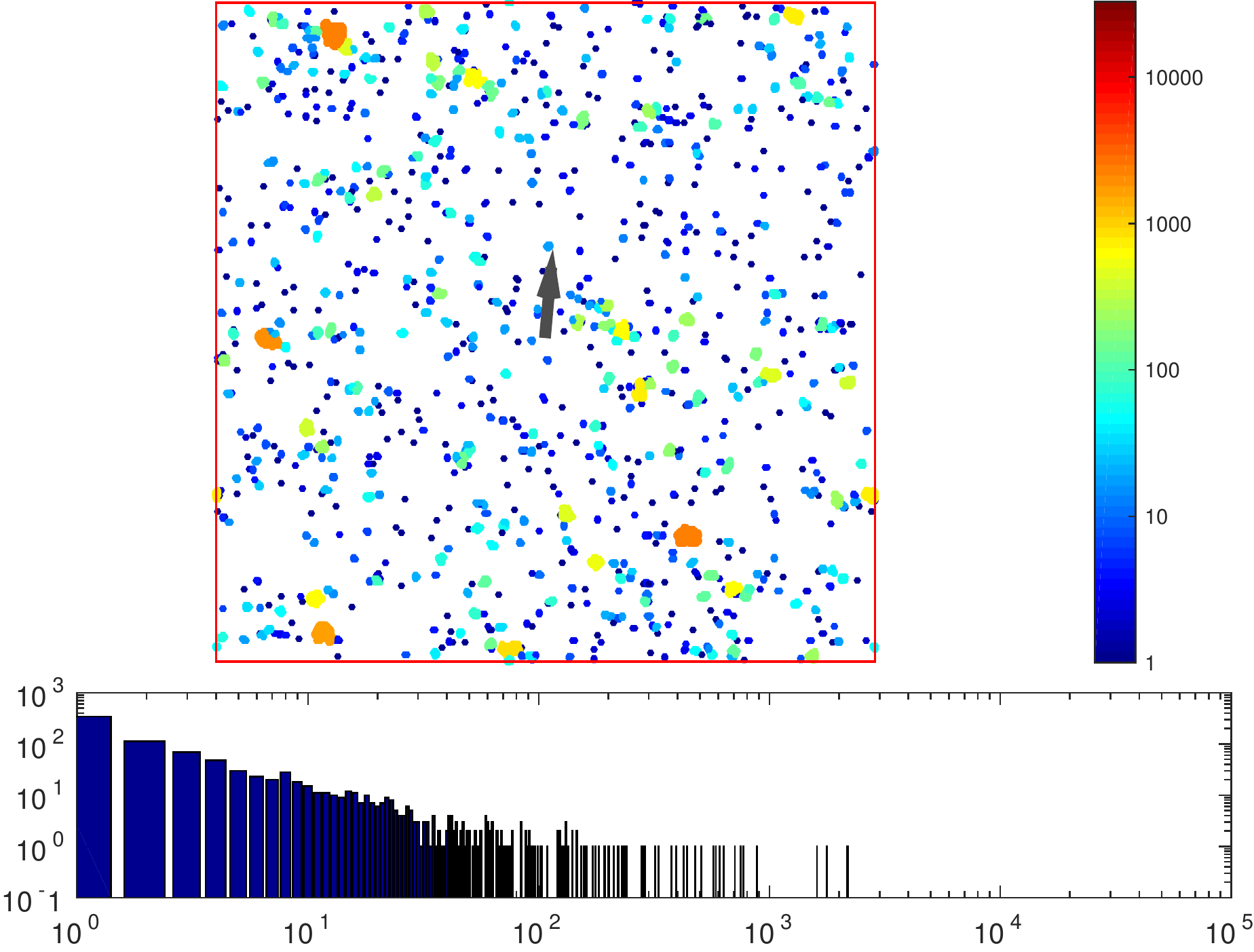}
   }
   \hfill
   \caption{Snapshots of configurations at different noise intensities. Black arrows show the average direction of motion. Lower panels show the CSD. (a) A single macrocluster is revealed. (b) Below and close to $\eta_{MC}$ the MC coexists with many single particles. (c) Above and close to the transition there are no intermediate size clusters. (d) A roughly spatially homogeneous \ffi\, distribution arises (see text). Parameters: $v_0=0.1$, $\rho=0.1$, $N=2^{15}$. Clusters are colored according to their sizes $m$. \ffi$>$0.98 for all situations. }\label{fig:conf}
\end{figure}
   
Because the system is out of equilibrium and lacks both Galilean invariance and momentum conservation, the nature of this ODT presents many features not observed in equilibrium systems. As an example the ODT in the original Vicsek model with metric interactions was first reported as a continuous transition \cite{Vicsek1995}. Later,  Chaté \etal  showed that it must be discontinuous, although the discontinuity could be  difficult to observe due to strong finite-size effects \cite{Gregoire2004, Chate2008}. Further work  provided new evidence on the first-order nature of this ODT, strongly unbalancing the scales \cite{Solon2015,Pattanayak2016}. On the other hand the topological counterpart of the model appears to be continuous \cite{Ginelli2010, Bhattacherjee2015} and curiously, finite size scaling made on systems where the discontinuos character of the ODT is not observable, both kinds of interactions give the same set of critical exponents  \cite{Baglietto2009,Barberis2014}. 
 
Besides, the ordered phase presents, at least close to the ODT, density waves transversal to the moving direction established by a symmetry breaking. Such waves are composed of clusters of many sizes that split,  merge and exchange particles. Thus, the identity of any cluster does not survive for a long time. Several authors have studied the clustering of this ordered phase, for example it is known that it has a \emph{cluster size distribution} (CSD) that follows a truncated power-law  \cite{Huepe2004} with an exponent  that diminishes when noise is lowered. Moving away from the ODT by further decrease of noise, the observed density waves seem to disappear \cite{Chate2008} and the global structure of the configurations become spatialy homogeneous. This last regime seems to be consistent with a ``homogeneous ordered phase'' theoretically predicted  by Toner and Tu \cite{Tu1998, Toner2012}. In fact, a binodal line in the noise-density phase space, that divides  this homogeneous ordered phase from the region were density waves are observed, was reported in the work of Solon \etal \cite{Solon2015}.  

The just mentioned power-law behavior was theoretically reproduced using a set of Smoluchowski equations by Peruani \etal \cite{Peruani2013}. In this last paper, a ``globally ordered'' phase at very low noise, different from a ``clustering'' phase, is mentioned but not analyzed in detail. Numerical solutions of the macroscopic equations show a peak in the CSD when the noise is low enough, suggesting the emergence of relatively large cluster. 
 
Thus, to much work was devoted to the understanding of the ODT and the ordered phase. As a consequence, there are few  results in the very low noise regime. Many articles alluded to the existence of new phases or structural changes at low noise, but, to the author knowledge, no systematic analysis on their phenomenology has ever been reported.

In this work, the very low noise regime is carefully studied. The main result is the existence of a \emph{macrocluster} (MC) which contains (almost) all the particles in the system. A cluster is a set of connected particles and two particles are connected whenever they are separated a distance less or equal to $R_0$. The \emph{cluster size}, $m$, is the number of particles belonging to a given cluster. Thus the mentioned macrocluster is formed by a macroscopic fraction of the system particles. These particles are gathered in a compact region of the space that is incompatible with an homogeneous ordered phase as seen in Fig.\ref{fig:confMC}. In this way, starting with a noise close to that yielding the ODT transition, many clusters forming the known density waves appear (not shown). When the noise intensity is lowered, clusters are allowed to merge forming larger clusters but more homogeneously distributed in the space, Fig. \ref{fig:confHomo}. Further decrease of the noise allow clusters to merge in larger ones, Fig. \ref{fig:confTFH}.  This phenomenon closely resembles the coalescence of clusters in equilibrium systems, with the difference that there the clusters move diffusively and here the clusters move ballistically because the persistence length seems to be larger than the mean free path. These large clusters share the space with small ones and the size difference increases as the noise intensity is lowered as seen in Fig. \ref{fig:confTF}. After further noise lowering, a macroscopic fraction of the particles merges into the mentioned single localized cluster shown in Fig. \ref{fig:confMC}.  In all panels of Fig. \ref{fig:conf}, clusters were colored according to their size, i.e. the number of particles that belong to them, and the corresponding  distribution of cluster sizes is depicted below each one. 

As it will be discussed, the expressions  ``global order phase'', ``polar liquid phase'', and ``homogeneous ordered phase'' are not compatible with the MC emergence.  This phenomenon appears at a noise intensity below that the investigated in most previous research, i.g. \cite{Peruani2013,Solon2015}, and had been obscured by the nature of the particular macroscopic quantity used to describe the different phases: The ferromagnetic order parameter. Then, the term \emph{macrocluster regime} will be used to describe the noise region where the structure of the system exhibits the MC  and, \emph{clustered regime} will refer to the set of other ordered configurations, i.e. homogeneously distributed clusters or density waves. The expression \emph{macrocluster-clustered transition} (MCT) is also coined to refer to the location, in the parameters space, where change between the clustered and the macrocluster regimes  takes place. It is worth to mentioning that all of these structures are included in the ``ordered phase'' described by a high value of the ferromagnetic order parameter.

\section{Numerical results}\label{s:nuemrical}

Each particle in the system is assumed to follow the overdamped Vicsek Eqs. \eqref{eq:langevin}. The time step will be $\Delta t=1$ and the  interaction radius $R_0=1$ as in the original Vicsek model. Simulations to obtain time averages were performed for $N=2^{12}=4096$ particles because the computational effort at larger system sizes becomes prohibitive at very low noises. Periodic boundary conditions were implemented in order to easily compare the results presented here with those available in the literature. Simulations started with random distributed particles with random orientations, both instances of randomness being generated by white noise. The size of the system was properly set to obtain initial homogeneous densities $\rho=1$ and $\rho=0.1$. Single realizations, such as those used for the  snapshots among other, were carried out for $N=2^{15}=32768$ and are properly indicated wherever necessary.

\subsection{Time evolution}\label{ss:time}

Simulations started with all the particles placed at random positions and orientations. Then, the system is left to evolve under the action of the chosen noise intensity until it reaches a steady state as it will be discussed below.  

The time evolution of two cluster-related parameters is measured together with the ferromagnetic order parameter \ffi. One of these is the \emph{normalized number of clusters} $M^*=\left\langle M \right\rangle /N$, with $M$ being the number of clusters in the system at a given time and $\langle . \rangle$ the average over configurations. The other parameter is the  \emph{normalized average cluster size} \m$=\frac{1}{N} \langle \frac{1}{M}\sum_i^M m_i \rangle$, where $m_i$ defines the size of the $i^{th}$ cluster with $i$ labeling all clusters present in the system.

\begin{figure}[h]
 \centering
  \includegraphics[width=\textwidth]{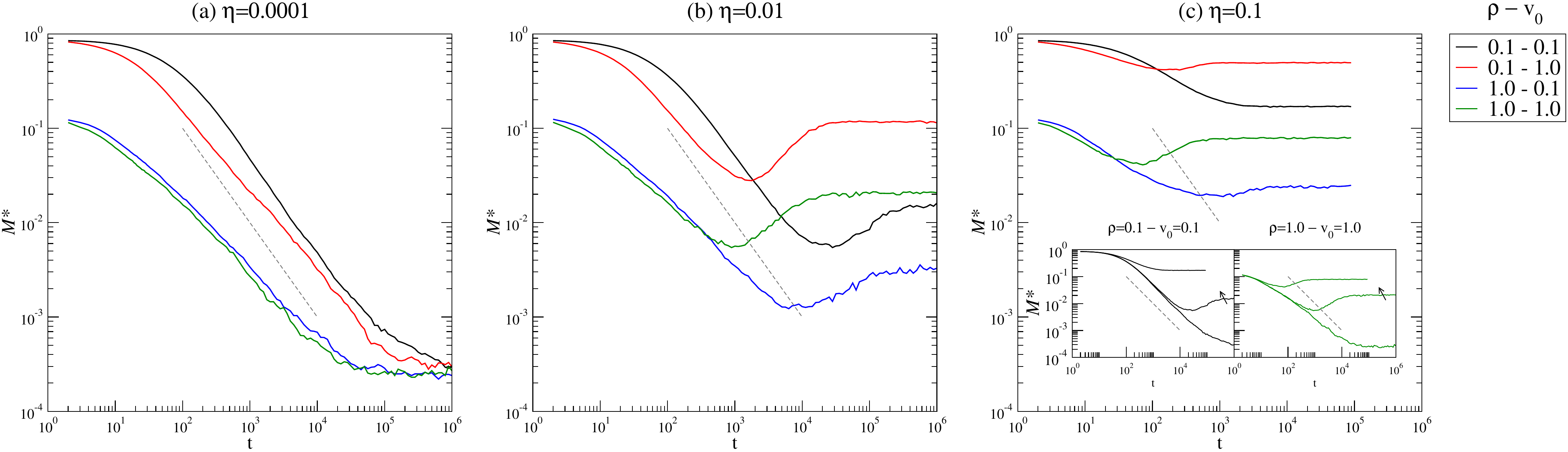}
   \caption{Time evolution of the mean number of clusters $M^*$ for different combinations of $\rho$ and $v_0$. (a) $\eta=0.0001$ where the macrocluster appears. (b) Clustered regime at $\eta=0.01$ in the homogeneous phase. (c) Clustered regime at $\eta=0.1$ where density waves are observed. Insets: the same information presented for various values of $(\rho,\, v_0)$ with the arrows illustrating the direction of increasing values of $\eta$. In all graphs the dashed gray line is a reference with slope -1. }\label{fig:M}
\end{figure}

The time evolution of the cluster size $M^*$ is shown in Fig. \ref{fig:M} where 30 realizations were averaged for four situations describing dense and dilute systems at high and low speeds for three noise values. The first interesting result is that, at noise intensities corresponding to the clustered regime, i.e. not very low ones, \M\, is a non-monotonic function of time exhibiting an extremum before it stabilizes, Fig. \ref{fig:M} panels (b) and (c). The minimum in the curves occurs because, at the beginning, the incipient clusters are able to collide from all possible directions. 
Those collisions merge them, in a sort of active coalescence, resulting in larger and larger clusters. This mechanism  leads to the establishment of the \emph{polar order} defined by a value of \ffi\, close to one.
Thus, in this stage, the collision rate, wich depends on a scattering cross section, must be larger than the fragmentation rate of clusters, wich depends on noise intensity \cite{Peruani2013}. 
When the polar order is established, the system exhibits an incipient density wave and, because clusters move almost parallel to each other, their cross sections diminish and so does the collision rate. 
At this point fragmentation starts to dominate the behavior of the system raising the value of \M. This increment in the  number of clusters will increase again the number of collisions until both, collision and fragmentation rates, become balanced at the final steady state. 
Note that the mentioned density wave, formed at the minimum of \M\,, will become disrupted at $\eta=0.01$, leading to the homogeneous ordered phase, cf. Fig. \ref{fig:confHomo}. 

In the low density regime and, because of the strong noise, the  information about the alignment of particles takes more time to spread over the system. Thus \ffi\, and \M\, reach the steady state at almost the same time leading to a monotonic decrease of the average cluster number as shown by the black line in  Fig. \ref{fig:M}(c). The percentages of the particles involved in the largest cluster at low densities are 0.050(5)\% and 0.010(7)\% for low and high speed respectively, i.e. even the largest cluster is indeed small. Since the life span of these clusters is quite short, the coarsening-like process is not possible.

On the other hand, in panel (a) that corresponds to the macrocluster regime, the curves are monotonically decreasing functions of time and they reach the same final steady state $M^*\simeq1/N$, i.e., there is a MC. Because the low noise, fragmentation rate is still smaller than collision rate even after the polar order is established. 

An interesting feature is that there is a region where it is possible to fit $M^*$ to a power-law $M^*=at^{-\gamma_M}$. For the dilute and low speed regime, black line in panel (a), such a behavior appears after a long first transient where nucleation takes place and an exponent $\gamma_M \simeq1$ is found. Thereby clusters grow as in a sort of active coarsening of droplets similar to the described in the classical work of Lifshitz, Slyozov and Wagner (LSW) \cite{oswald}. Active coarsening just means that small clusters not last for long time because they are captured by larger ones. But, unlike Ostwald ripening, active clusters keep merging until just one remains.  
An increase of activity (speed) accelerates nucleation through an increase of the collision rate shortening the first transient as shown by the red lines in Fig. \ref{fig:M}. The coarsening-like process is indeed the active-coalescence of the clusters. As it was stressed by Tokuyama \etal \cite{Tokuyama1986} the LSW theory, that not include self propulsion of particles, holds only for low densities. 

On the other hand, at higher densities the incipient clusters form very rapidly, accelerating the coalescence because it is easier for particles to find a cluster and, as observed in Fig. \ref{fig:M}(a), the MC emerges at shorter times than in the dilute case. As a consequence, the measured exponents decrease in magnitude to a fitted value close to $\gamma_M\simeq3/4$. In such situation the density becomes a more relevant parameter than the speed.
Fitted values for $\gamma_M$ in the region where a power law behavior is observed is presented in Table \ref{t:M} for all curves in Fig. \ref{fig:M}(a).

 \begin{table}[h]
  \caption{ Power law exponent $\gamma_{M}$ fitted from curves in Fig. \ref{fig:M}(a). }\label{t:M}
 \begin{ruledtabular}
 \begin{tabular}{l|c|c|c|c}
  $\rho,\,v_0$  & \textcolor{black}{0.1 - 0.1} &\textcolor{red}{0.1 - 1.0} & \textcolor{blue}{1.0 - 0.1} &\textcolor{verde}{ 1.0 - 1.0}\\ \hline
&0.966(7)&0.846(9)& 0.738(7)&0.759(8) 
\end{tabular}
\end{ruledtabular}
\end{table}


In the insets of Fig. \ref{fig:M}(c), the same curves are now presented for the same system at different noise intensities. There, can be observed that the power-law behavior is cut-off as noise intensity increase. In the low density - low speed regime at the highest reported noise, such a cut-off is so strong that completely overwhelms coarsening as was mentioned before.

Recall that the values of the noise used in Fig. \ref{fig:M} belong to different regimes with different structural configurations of the system, all inside the ordered phase. The ferromagnetic order parameter can not take in to account this differences which only can be made evident by an order parameter related to the clustering dynamics, such as the cluster number \M\, or the mean cluster size \m\, used in this work.

\begin{figure}[h]
 \centering
  \subfloat[\label{fig:time-phi}]{
    \includegraphics[width=0.4\textwidth]{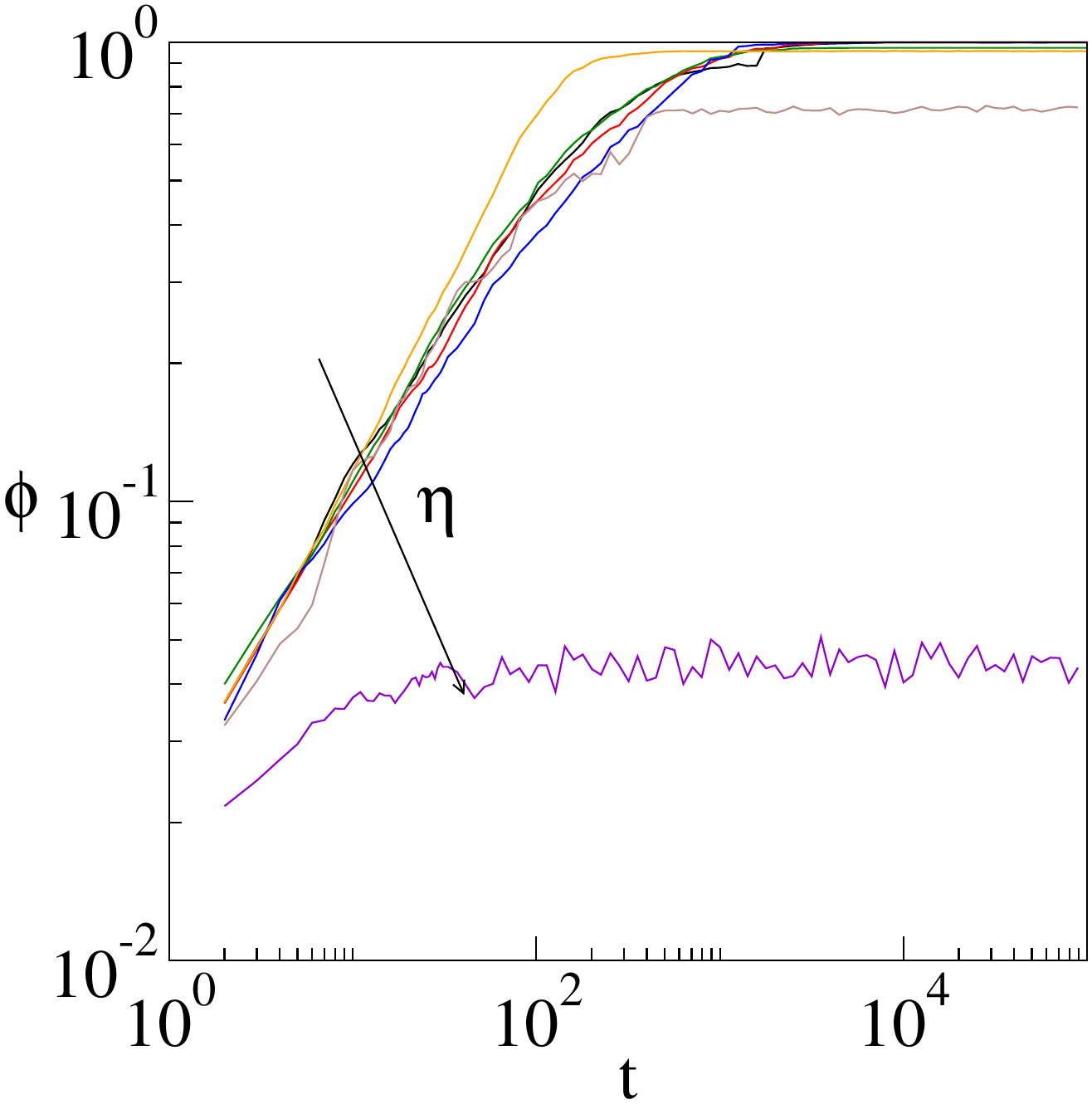} 
    }
    \hfill
   \subfloat[\label{fig:time-m}]{
   \includegraphics[width=0.4\textwidth]{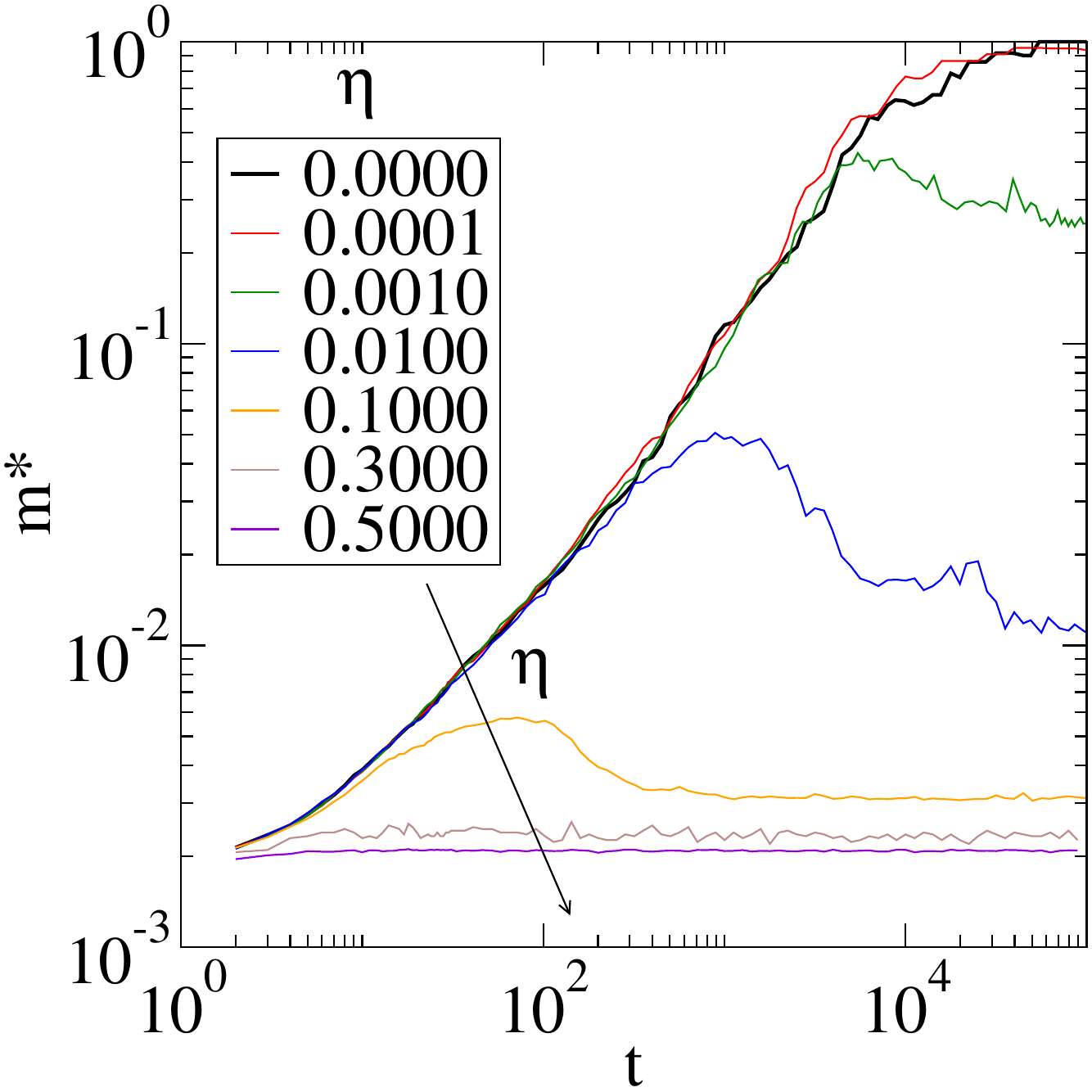}
   }
   \\
   \subfloat[$\rho=0.1$ - $v_0=0.1$ - $N=2^{15}$\label{fig:time-bump}]{
   \includegraphics[width=0.4\textwidth]{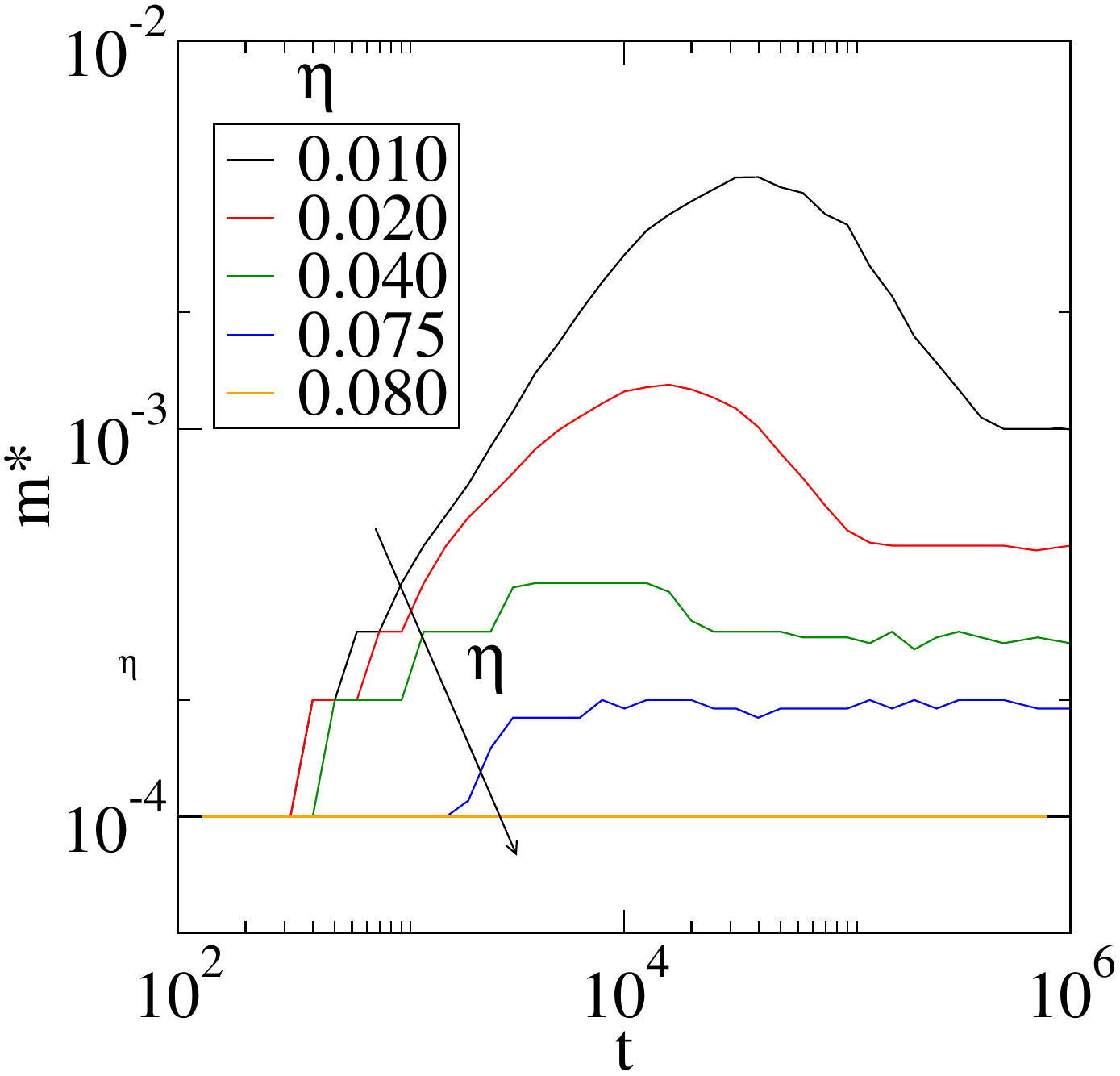}
   }
   \hfill
   \subfloat[\label{fig:time-zoom}]{
   \includegraphics[width=0.4\textwidth]{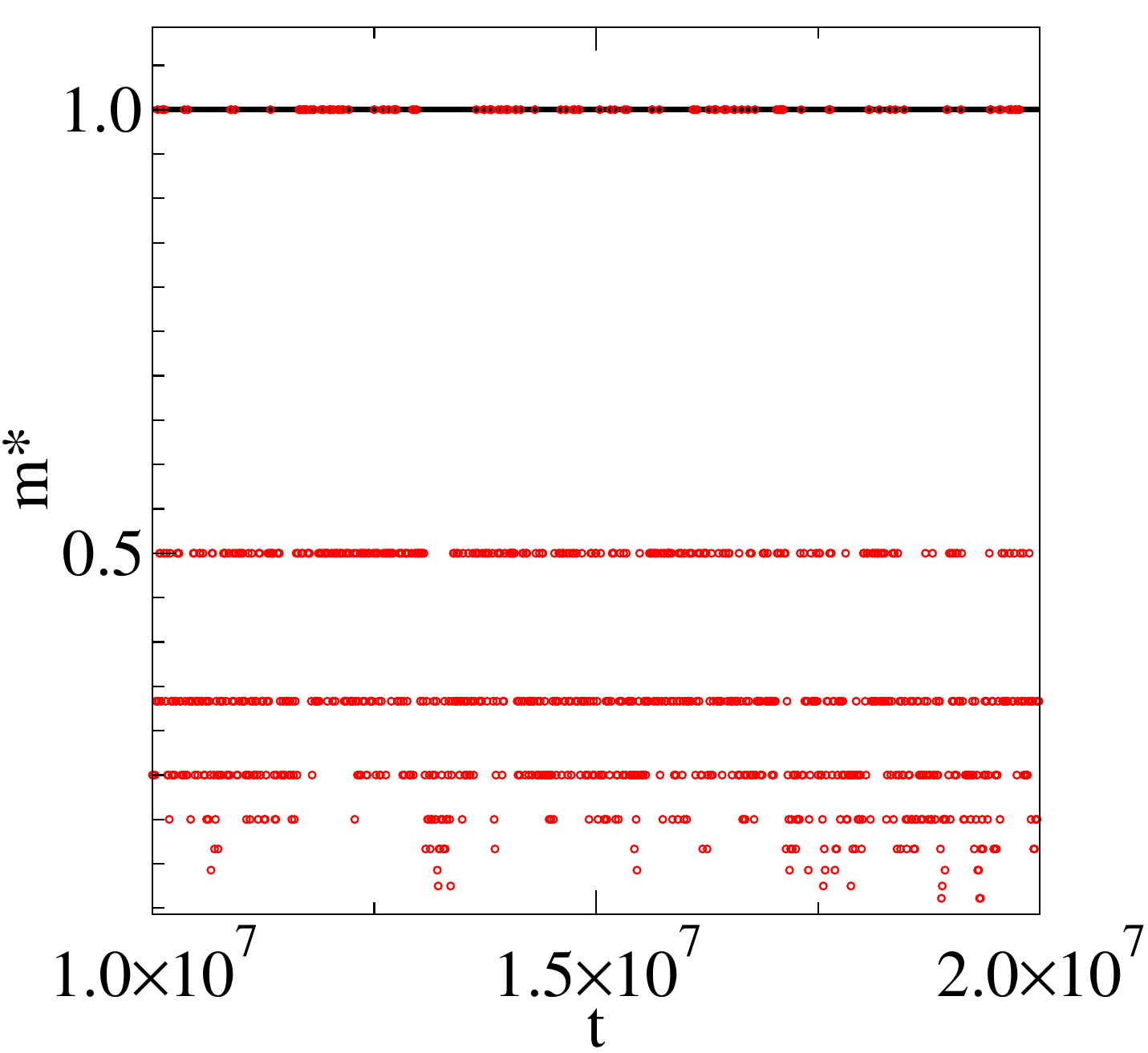}
   }
   \hfill
   \caption{Time evolution of  $\phi$ and \m. (a) The ferromagnetic order parameter, $\phi$, stabilizes over  $t\simeq10^3$. The curves show the same qualitative behavior for different noise intensities. (b) The average cluster size \m\, change its behavior at noise intensities close to $\eta\simeq 10^{-3}$ revealing the emergence of a macrocluster at lower noise intensities. Note the similitude between $\eta=10^{-4}$ (red) and $\eta=0$ (black) curves. (c) In the low density - low speed regime in a larger system ($N=2^{15}$), the flattening of the bump close to the ODT is most easily observed.  (d) Portion of one realization of the time series for \m, at times when the MC is losing and reabsorbing particles. The sampling rate is $5\times 10^4$ time steps.  }\label{fig:time}
\end{figure}

To illustrate this last point, the time evolution of \ffi\, and \m\, is depicted in Fig. \ref{fig:time} panels (a) and (b) respectively. Comparison between these functions shows that, at low noise, the time taken to reach the steady state is lower for \ffi\, than for \m\,. Note that \ffi, that describes the evolution of the alignment process, always stabilizes more or less at $10^3$ time steps for the smaller noise intensities. In particular, the time taken to reach the fully ordered  state ($\phi\simeq1$) diminishes  when noise intensity is raised. This can be observed in panel (a) where the orange line ($\eta=0.1$) grows faster than the black ($\eta=0$) to blue ($\eta=0.01$) lines. Note that the orange curve corresponds to a noise intensity well inside the clustered regime where density waves are observed. At higher noise intensities, close to the ODT, order decreases (brown curve) and the steady state of both \ffi\, and \m\, is again quickly achieved. Beyond the ODT noise intensity the system stays disordered for any time (violet curve).

The minimum shown in the time evolution of \M\, corresponds to a maximum in the time evolution of \m and, in the following, both will be called \emph{extremes}. 
The steady state of \m, takes significant different time being most delayed for most lower noises. This effect is more evident in panel (c),  where the time evolution of \m\, is depicted for a larger system ($N=2^{15}$) in the low density - low speed ($\rho=0.1$, $v_0=0.1$) regime. 
The ODT in this case occurs at $\eta\simeq0.75$ (blue line); the amplitude of the extreme becomes larger and its position shifts to the right when the noise intensity is decreased. Beyond the ODT the disorder prevails at all times (orange curve). Thus, the extreme flattens close to the ODT and beyond when noise always overcomes alignment forces. 
Therefore, close to the ODT, both \ffi\, and \m\, stabilize almost at the same time because active coalescence occurs between transitory clusters. The extreme also disappears at the MCT noise and below becoming a feature of the clustered regime only consequence of the late balance between self-diffusion, self-propulsion and alignment forces. The extremes become an important effect to consider when cluster-related quantities, as the average neighbor number, will be measured.

At noise intensities below $\eta=0.001$, green line in Fig. \ref{fig:time-m}, the value of \m\, increases non-monotonically until $m^*$ reaches a value very close to one, meaning that there is a single cluster most the time. However, the MC will lose a few particles in time, that will come back to hit it due to periodic boundary conditions. 
This losing/gaining process can be observed in the behavior of \m\,  depicted in Fig. \ref{fig:time-zoom} for one realization at $\eta=0.0001$. The red curve was sampled every $5\times10^4$ time-steps and shows that some particles abandon the MC to later be reabsorbed again. Note that, in the example, \m=0.5 means the average of one cluster of size N-1 plus one of size 1 normalized by $N$. That is why the  average on realizations shown in (b) gives values of \m\, a little smaller than 1. Because at zero noise there is no source of fluctuations, it must be an absorbing state: Once the MC  is formed, there are no perturbations that change its configuration. It is the case  depicted by the thick black line showing that the MC remains unmodified for all time.


\subsection{Cluster size distribution}

As it was mentioned before and illustrated in Fig. \ref{fig:conf}, a clear MC is observed if $\eta\lesssim0.001$. Note that the cluster size distribution (CSD) shown in panel (a) is just a peak  at $m=32768$, the system size.  Close to the noise intensity that marks the onset to the MC regime, panel (b) with $\eta=0.001$, there are very large and very small clusters in an unsettled coexistence. Indeed, clusters appear colored in grades of red and blue without the presence of yellow and green clusters and the cluster size distribution is void at intermediated sizes indicating that the probability of finding intermediate cluster sizes is very low.

Increasing the noise intensity to $\eta=0.002$, panel (c), intermediate cluster sizes become possible. At last, in panel (d), the noise is high enough ($\eta=0.01$) to strongly decrease the probability of obtaining very large clusters, but still low enough for the system to be well inside the ordered phase ($\phi\simeq 1$). Clusters whose size is close to the size of the system become infrequent and they last few time steps. 

\begin{figure}[h]
 \centering
  \includegraphics[width=0.8\textwidth]{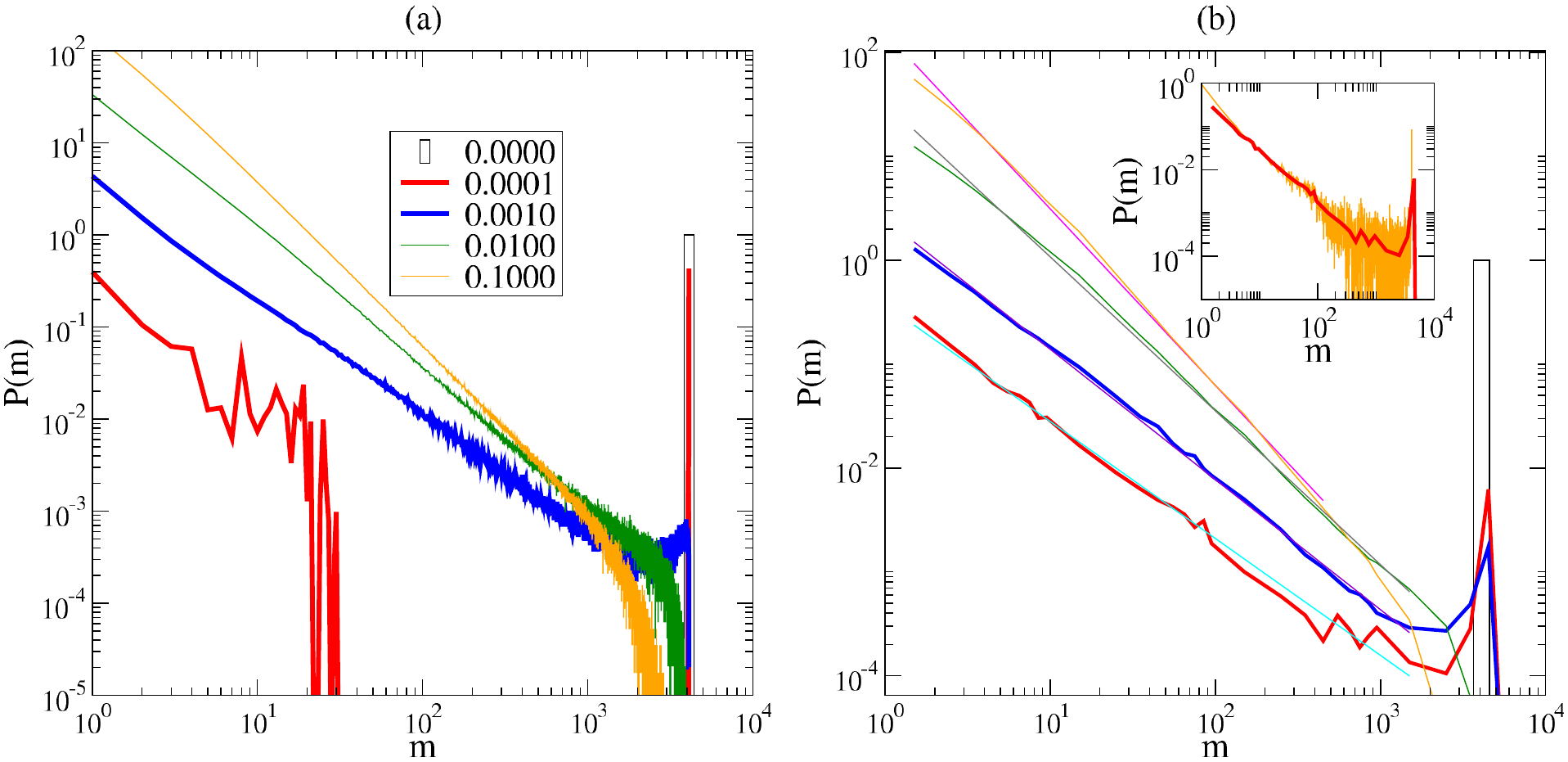}
   \caption{Cluster size distributions $P(m)$ averaged over $10^4$ realizations. (a) CSD with linear binning. (b) CSD with logarithmic binning clearly reveals that at very low noise, $\eta=10^{-4}$ (red), the MC ($m\simeq N$) exhibits a process of particle gain/loss. As the noise is increased, the well known power law plus a cutoff behavior emerges. Note that close to the transition,  $\eta=10^{-3}$ (blue), the CSD becomes non monotonic. Straight lines correspond to power law fittings. Inset: CSD with linear (orange) and logarithmic (red) binning at $\eta=10^{-4}$ averaging $10^5$ realizations (see text).   }\label{fig:CSD}
\end{figure}

To quantify and better understand this feature, clusters size distributions  were measured for the situations presented in Fig. \ref{fig:conf}. To do this $10^5$ CSD, as those depicted in Fig. \ref{fig:conf}, were obtained from  configurations.  They were collected every $10^3$ time steps in the steady state and then, averaged to obtain a distribution $P(m)$. The behavior of $P(m)$, for five noise intensities, using linear and logarithmic binning is respectively shown in panels (a) and (b) of Fig. \ref{fig:CSD}. Note that, in panel (a), averaging over $10^4$ realizations is not enough to obtain a regular shape of the curve for the MC regime (red). Thus, in panel (b) the red curve was instead averaged over $10^5$ realizations to obtain a better description of its behavior. In the inset of panel (b), linear (grey) and logarithmic (red) binning are shown for the CSD at $\eta=10^{-4}$ were the MC appears. Thus, in panel (b), this last CSD is depicted instead  of the original one.

The transition between both regimes becomes now evident: For high noise intensities (yellow and green), $P(m)$ is a continuously decreasing power-law distribution plus a finite size cutoff. In the opposite case, for low noise intensities, there is a drastic change for $\eta=10^{-3}$, where $P(m)$ becomes non monotonic and shows a peak at large cluster sizes pointing to the emergence of an MC (blue and red). Furthermore, the shape of  $P(m)$ shows that the MC is not always formed by all particles at non-zero noise but sometimes it will lose particles that will form transitory clusters. In simulation runs with sizes from $N=2^{10}$ up to $N=2^{15}$, the MC has always lost some particles over time. The clusters described by the left part of the distribution correspond to those single particles that have interacted with other loose particles before being reabsorbed by the MC. Periodic boundary conditions give to released particles a high chance of hitting the MC again allowing their reabsorption. Large clusters could appear but they are very rare,  the only observed instances occurring in few realizations. An example is when the  system form two large clusters that gather the particles, containing more or less half the particles each. When finally they collide to form the MC, two situations may occur: if the colliding particles velocities are  roughly parallel, the resulting MC has two high density nuclei and fluctuations can split it again. The second situation, the most common one, occurs when those two large clusters collide and merge into a mono-nucleus MC.

The blue line in Fig. \ref{fig:CSD}(b), for $\eta=0.001$ close to the onset to the MC regime,  shows that the probability of finding intermediate cluster sizes is really low but still nonzero.  The distribution clearly follows a power law that does not cutoff, but raises to a peak for clusters with size $m\simeq N$, i.e. the size of the system. These larger clusters last a short period of time after which they split again. As it will be shown below, a phase transition takes place for a noise intensity $\eta\gtrsim 0.002$, for which the cutoff reaches the system size because large fluctuations allow for the presence of unstable large clusters.  These results are consistent with those obtained by, among others, Huepe and Aldana \cite{Huepe2004}. In particular, these authors mention that there is a power law with a cutoff close to the ODT but leave the study at lower noises for further work. They also mention that the exponent $\gamma(\eta)$ of the power law $P(m)=a\,m^{-\gamma(\eta)}$ decreases with noise, a result confirmed here by the measured value for the exponents $\gamma(0.1)=1.70(2),\,\,\gamma(0.01)= 1.48(1), \,\, \gamma(0.001)=1.254(8)$ and $\gamma(0.0001)=1.13(2)$, obtained by fitting the left part of the  distributions.
The peak close to the transition was described theoretically by Peruani and Bär \cite{Peruani2013} but they do not do further analysis and no CSD consistent with the macrocluster regime was reported.  On the other hand, this model was used by  Starruss \etal \cite{Starruss2012} to describe myxobacteria mutants able to move just forward. There, it is found that ``At low and intermediate densities, non-reversing cells display collective motion in the form of large moving clusters'', with a critical density above which clusters can be arbitrarily large. The CSD shown here describes this same result.

Finally it is interesting to compare these results with those published by Chaté \etal \cite{Chate2008} p.10 and Fig. 15. In that work it is mentioned that ``at low noise bands stand less sharply out of the disordered background'' until they ``abruptly disappear and are no longer well-defined transversal objects''. It is concluded that the ``the local order parameter is strongly homogeneous in space''. This last observation is qualitatively verified  by inspection of Fig. \ref{fig:conf}(d). Indeed, these authors were very close to see the MCT, but they did not  consider low enough noise intensities. As a consequence, the drop shown in their Fig. 15 does not correspond to the transitions between regimes but  just to the dilution of the band as they stated. Such a drop is related to the noise range where the CSD begins to become bimodal.

\subsection{Phase portrait}\label{s:phase}

\begin{figure}[h]
\subfloat[Configurations\label{fig:r1-conf}]{
    \includegraphics[width=0.48\textwidth]{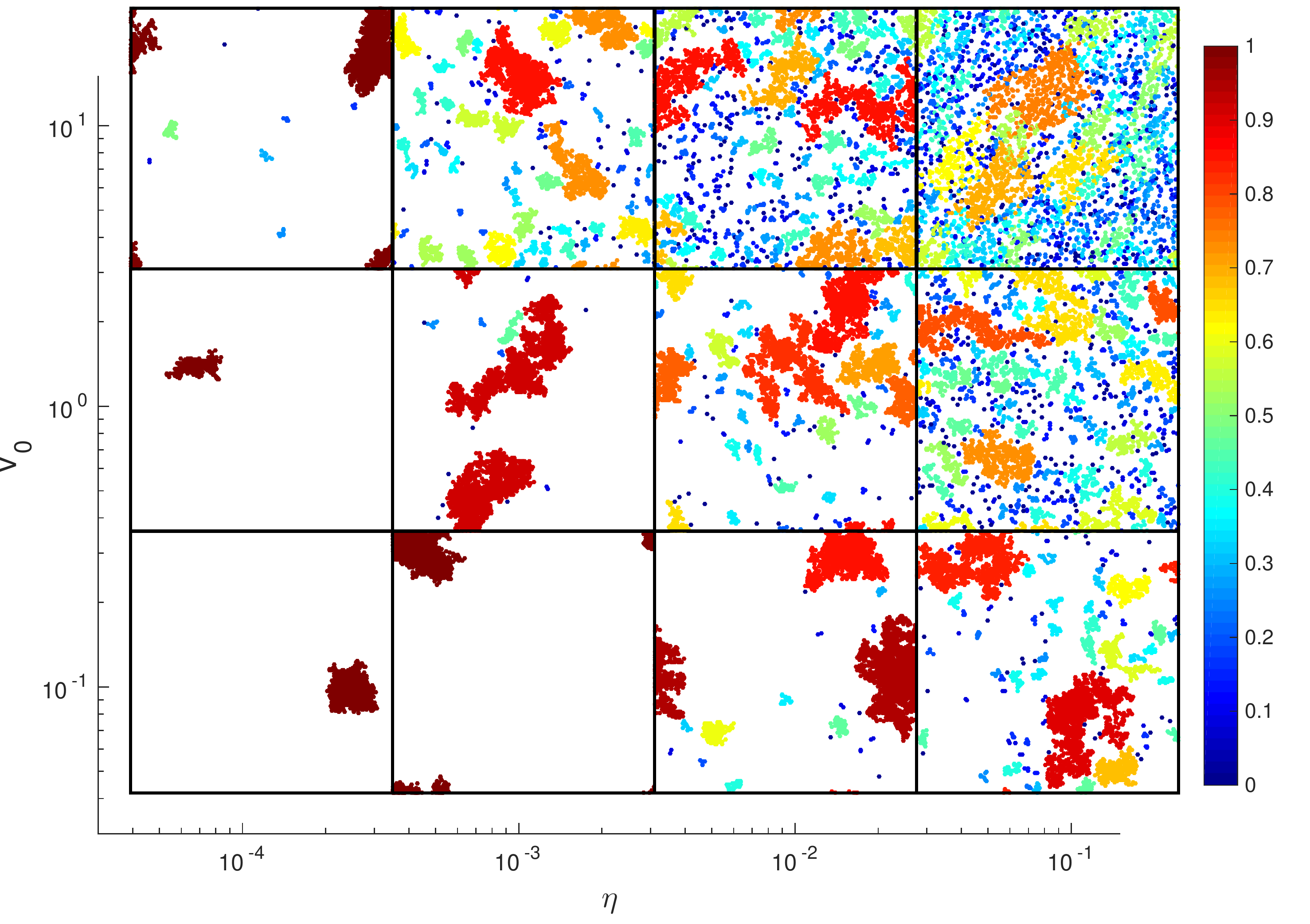} 
    }
   \subfloat[\m\label{fig:r1-mmedio}]{
   \includegraphics[width=0.48\textwidth]{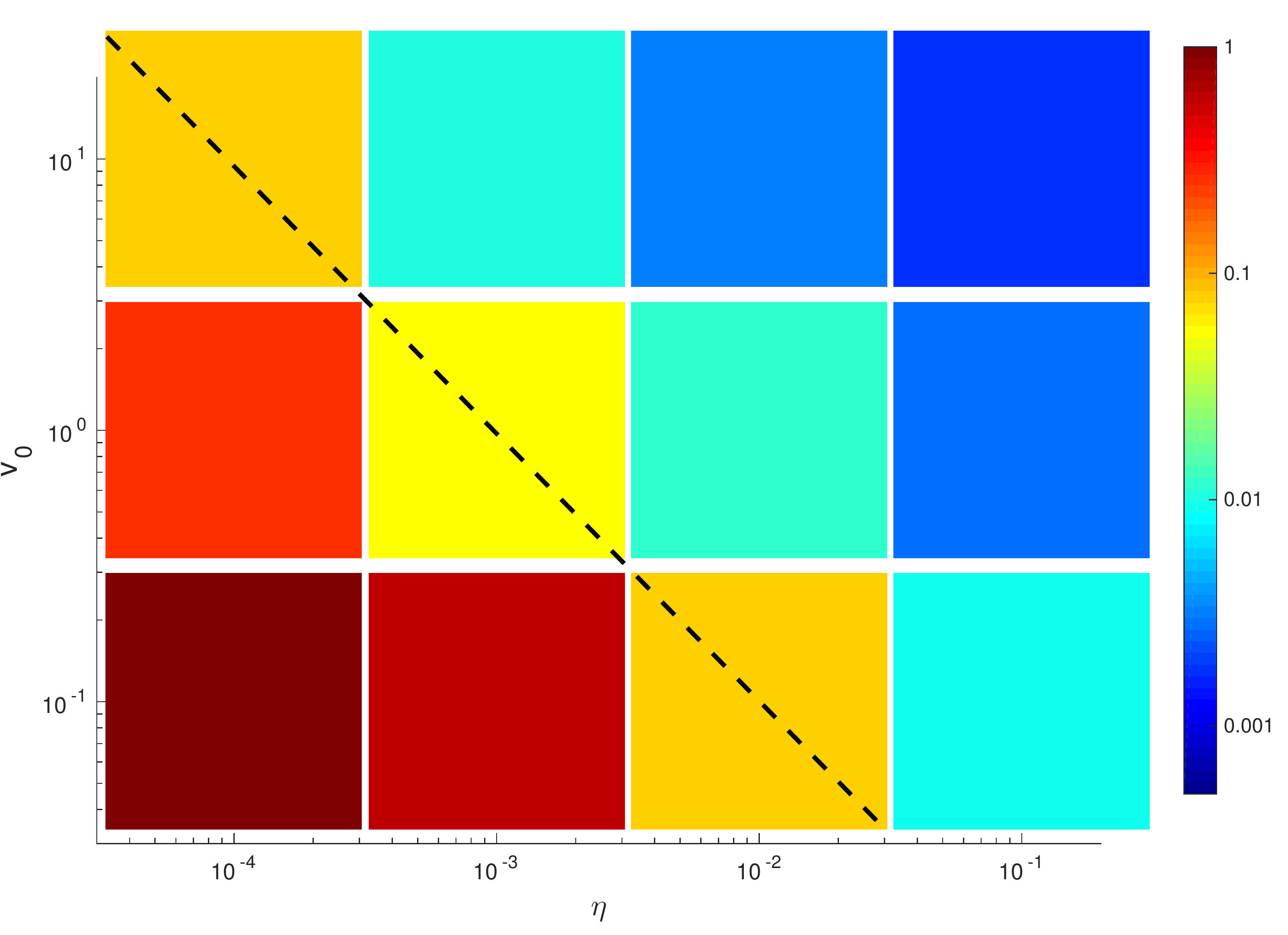}
   }
 
\caption{Phase portrait $v_0$-$\eta$ for $\rho=1$. Dashed line is a rough estimate of the transition line between  clustered and macrocluster regimes. } \label{fig:phase1} 
\end{figure}

Here we analyze the influence of two free parameters, $v_0$ and $\rho$, is here analyzed. To do this, simulations with $N=2^{12}$, changing both the speed $v_0$ and the density $\rho$, were carried out. In Fig. \ref{fig:r1-conf}, configurations for $\rho=1$ are shown in the $v_0-\eta$ phase portrait. Complementary, Fig. \ref{fig:r1-mmedio} shows the corresponding values of \m, which are the time averages over 100 realizations, sampled every $5\times10^4$ time steps. Note that in Fig. \ref{fig:r1-mmedio}  the color scale was set to be logarithmic, because \m\, falls arithmetically with particle loss as was discussed in Sec. \ref{ss:time}. At the lower left corner of these figures, the macrocluster regime can be clearly identified in both, configurations and \m$\simeq 1$. In the upper left corner of Fig. \ref{fig:r1-conf}, a big cluster in dark red coexists with some small clusters. Looking at the corresponding \m\, value, it becomes clear that this cluster is incidental and will not last long enough as to be considered an MC. Hence the parameter \m\, clearly discriminates the region where the MC stands (in dark red) and where there is a monotonic distribution of clusters (blue region). The yellow band roughly defines a macrocluster-clustered \emph{transition line} where the above mentioned coexistence among big and small clusters lies. An estimate of its position was drawn as a dashed line in the  graph.

\begin{figure}[h]
\subfloat[Configurations\label{fig:r01-conf}]{
    \includegraphics[width=0.48\textwidth]{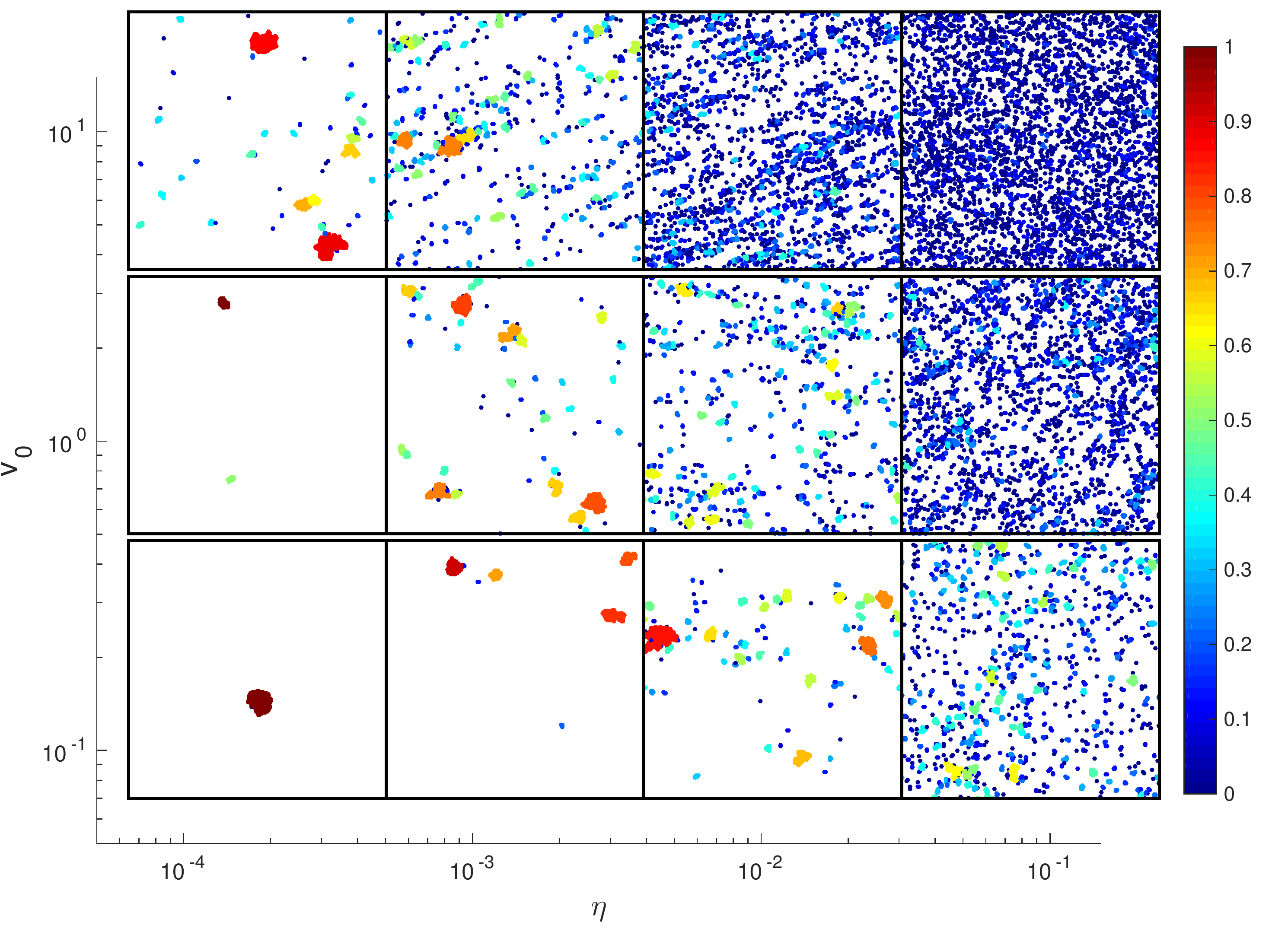} 
    }    
   \subfloat[\m\label{fig:r01-mmedio}]{
   \includegraphics[width=0.48\textwidth]{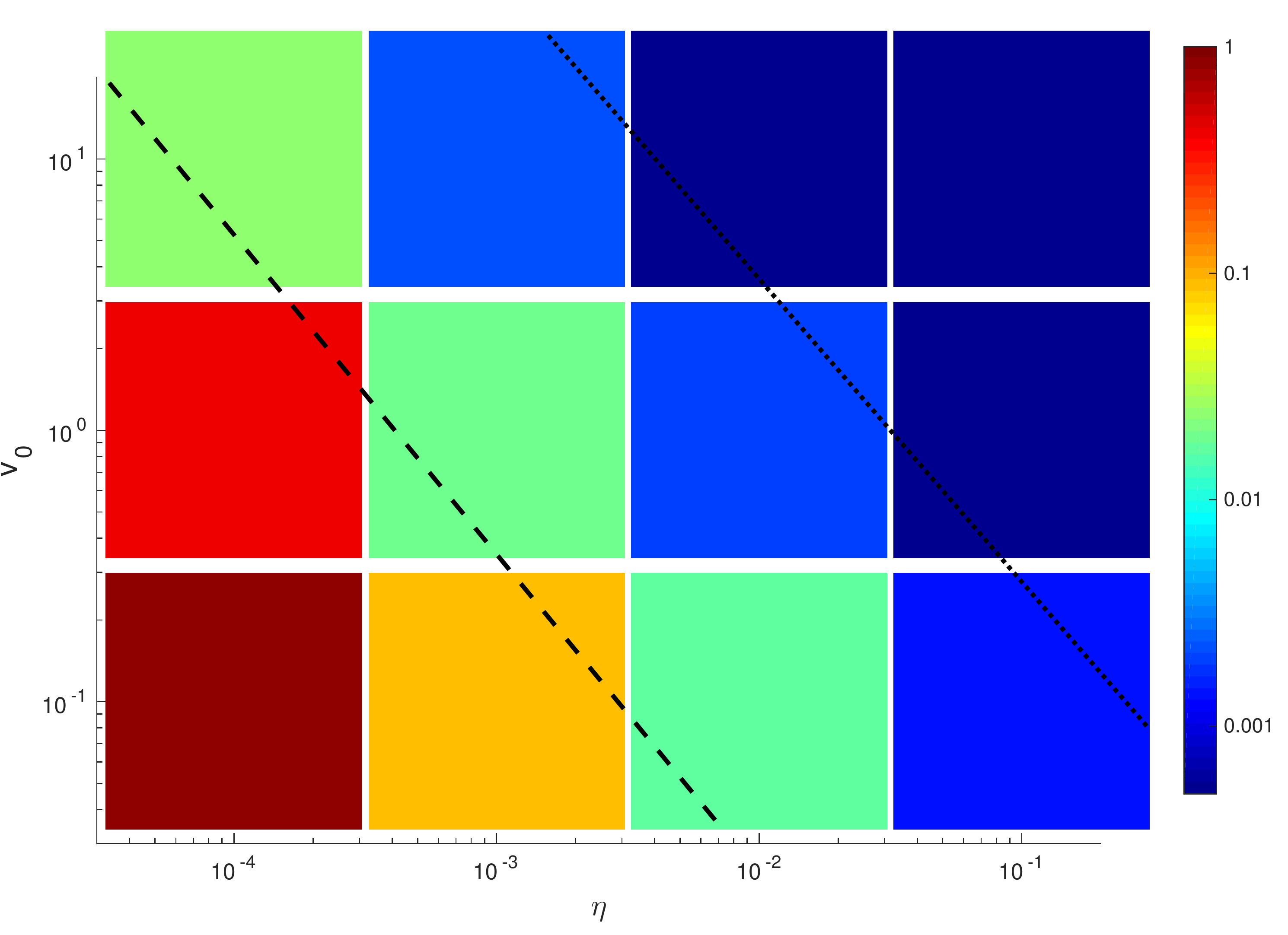}
   }
 
\caption{Phase portrait  $v_0$-$\eta$ of configurations for $\rho=0.1$. It looks similar to $\rho=1$ with a presumably steeper transition. Dashed and dotted lines are estimates of the MCT and ODT lines respectively.  } \label{fig:phase01}
\end{figure}

The behavior of systems at lower density ($\rho=0.1$) is presented in  Fig. \ref{fig:phase01}. Panel (a) shows configurations in the steady state where a qualitative shift to low noise intensities of the MCT is appreciated. In the phase portrait for \m (b)  there is a green fringe that lies where the transition line was observed for a higher density in Fig. \ref{fig:r1-mmedio}. The evidence is not strong enough as to determine if there is a change in the slope of the estimated transition (dashed) line  or if it becomes non-linear. However a narrowing of the transition zone seems the most plausible explanation. Note that the gas-like (disordered) region appears where an ODT (dotted) line was approximately drawn. As it was pointed out in \cite{Gregoire2004}, finite size-effects are somewhat weaker at lower densities. So the low-speed, low-density regime is a good candidate to study the transition in detail. 

At higher densities (i.e. $\rho=10$)  the maximum packing fraction $\Phi=\frac{2\pi R_0}{L^2}$ is really high. Hence, it becomes easier for the  particles to interact with each other and to form  just one cluster. The study of such a regime is out of the scope of this paper.

To sum up, this is a new piece of the phase portrait presented by Solon \etal \cite{Solon2015}. Note that the lowest noise intensity in that work was $\eta=0.1$, which coincides with the highest intensity considered here. As a consequence the results  shown in the present work broaden the available phenomenological description of the Vicsek model.

\subsection{Finite size scaling}

To complete the picture, a finite size scaling analysis was performed to characterize the MC regime as a phase transition. In Sec. \ref{s:phase} it  was conjectured that the low density - low speed regime presents a narrower transition region than the high density - high speed regime. For this reason, the scaling analysis was carried out for $\rho=0.1$ and $v=0.1$. 

Because the focus is on the cluster distributions, two quantities borrowed from percolation theory were considered:    The sizes of the first $\zeta_1=\max_{i=1,N}\langle m_i \rangle$ and $\zeta_2$ (similarly defined) of the second largest cluster. They were measured in CDS's as functions of the noise intensity $\eta$ (in this section $\langle \cdot \rangle$ will represent the  average over all collected data, i.e. summing up over time and realizations.)  The chosen system sizes were   $N=16584,\,8192,\,4096,\,2048,\,1024,$ and $\,512$ particles. Usually scaling is performed on the lenght $L$ of the supporting space. However, since the initial density $\rho=N/L^2$ remains constant and, in the MC regime, the single cluster is compact, the particle number $N$ becomes a natural system size variable. 

Up to 30 realizations of the system evolution for each value of the noise and the system size were ran. Once the steady state of each realization was reached, 1000 data from each realization, i.e. the size of  first and second largest clusters, were collected at a sampling rate of $10^4$ time steps for systems with $16584$ and $8192$ particles and, $10^3$ time steps for systems with $4096,\,2048$ and $\,1024$ particles. 

\begin{figure}[h]
    \includegraphics[width=0.9\textwidth]{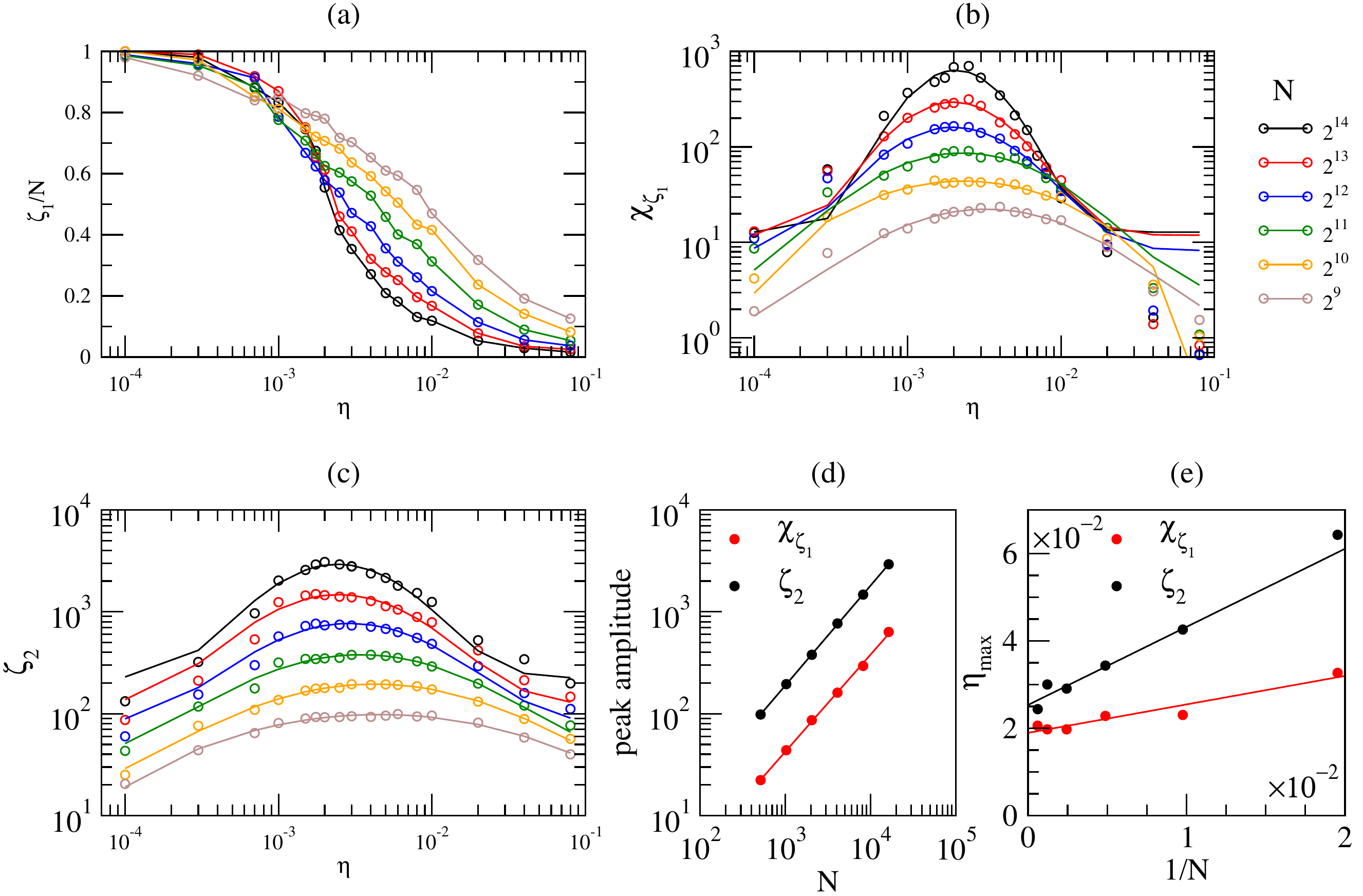} 
\caption{ Transition to the emergence of an MC. (a) The averaged largest cluster size $\zeta_1/N$ falls from 1 at nonzero low noise intensity.  Fluctuations of $\zeta_1$ (b) and $\zeta_2$ (c) as described by the corresponding susceptibilities, wich develops peaks close to the critical noise. Lines are lognormal fittings. (d) Fitting of the scaling amplitudes gives the trivial power-law b $\chi(\zeta_1)\sim N $.  (e) The peak position $\eta_{max}= \eta_c + const \frac{1}{N}$. The extrapolation to $N\rightarrow \infty$ gives $\eta_c=0.00183(8)$ and  $\eta_c=0.0025(5)$ for the largest and second largest clusters, respectively.  } \label{fig:scaling}
\end{figure}

The dependence on $\eta$ and $N$  of the normalized size of the  largest cluster, $\zeta_1/N$, is depicted in Fig. \ref{fig:scaling}(a). Note that the scale in the abscissa is logarithmic for a better analysis. The curves show that the largest cluster becomes macroscopic for very low, \emph{but non zero} noise intensity and that its size decays more steeply with  $\eta$ as $N$ increases. Thus, effectively there is a transition from a single cluster with the size of the system, $\zeta_1/N\simeq1$, to CSD's where large clusters become rarer and rarer as the noise intensity increases. 

The signature of a phase transition is the presence of large fluctuations, close to the critical noise $\eta_c$, of a suitable order parameter. This signature emerges clearly in Fig. \ref{fig:scaling}(b), that shows the susceptibility of $\zeta_1$, defined as $\chi_{\zeta_1}=\langle \zeta_1^2 \rangle- \langle \zeta_1 \rangle^2$, versus noise intensity. Note the peak  close to the critical noise, whose maximum amplitude scales with the following scaling ansatz:   
\begin{equation}\label{eq:scaligchi}
 \chi_{\zeta_1} \sim N^{\xi_1},
\end{equation} 

\noindent where $\xi_1$ is a critical exponent. 

The size of the second largest cluster, depicted in Fig. \ref{fig:scaling}(c),  also develops the expected peak close to the critical noise. Its maximum satisfy the scaling law

\begin{equation}
 \zeta_2 \sim N^{\xi_2},
\end{equation}

\noindent with $\xi_2$ being another critical exponent.  

The curves in Figs. \ref{fig:scaling}(c) and (d) were fitted by lognormal functions to obtain an approximation of the position and amplitude of the peaks. In Fig. \ref{fig:scaling}(d) the scaling of both amplitudes are shown to scale in a trivial way with exponents close to one: $\xi_1=1.05(6)$ and $\xi_2=0.95(6)$. 

It is also expected that the critical threshold $\eta_c$ for the MC emergence will depend with the system size. In Fig. \ref{fig:scaling}(e), the position of the peaks as functions of $1/N$ are depicted to show that they obey 
\begin{eqnarray}\label{eq:chi}
 \chi_{\zeta_1} = \eta_c + a \frac{1}{N},\\
 \zeta_2 = \eta_c + b \frac{1}{N},
\end{eqnarray}

\noindent with $a,b$ suitable constants.

The obtained values $\eta_c=0,00189(8)$ and $\eta_c=0.0025(5)$ are statistically indistinguishable and become good approximations for the magnitude of the critical noise in the thermodynamic limit.

Finally it is worth to mentioning that the chosen variables are not able to detect the transition between the homogeneous ordered and the density waves phases. Because the CSD's in those phases show exponential cut-offs that regularly decrease with noise intensity, the largest clusters sizes are always bounded.

\section{Discussion} \label{s:discussion}
  
There have been various  attempts to describe the simulation results of Vicsek's model  in terms of field equations. As an example, as it was stressed in \cite{Chate2008}, the discontinuous nature of the phase transition was first theoretically proposed but not observed due the mentioned strong finite size effects. On the other hand, Baglietto \etal \cite{Baglietto2013} showed that the bands observed in the ordered phase are only a consequence of the boundary conditions. This last feature seems to run against theoretical predictions \cite{Tu1998,Guttenberg2014} and, to the author knowledge, has not yet been refuted. 
  
Hence, the meaning and properties of the order-disorder phase transition, that Solon \etal propose as a liquid-gas one \cite{Solon2015}, remain an important discussion point. For this reason, the nature of the ODT in Vicsek models is, even after more than twenty years,  a strong motivation for active research in both approaches:  numerical simulations and field equations. Some relevant simulation results were discussed in Section \ref{s:intro}. Here a brief discussion of the  theoretical approaches is presented.      

The first attempts to obtain field equations were those due to Toner, Tu and coworkers \cite{Toner1995,Tu1998,Toner2012}  where phenomenological hydrodynamic equations were proposed to describe the symmetry breaking that favors the ordered state. Furthermore, critical exponents were obtained \cite{Toner1998,Toner2012} although later Grégoire \etal \cite{Gregoire2004} wrote that the estimations were ``loose''. But, most important, these works described the existence of density waves and homogeneous order. In this line Guttenberg \etal \cite{Guttenberg2014} showed that a quasi one-dimensional density wave could appear independently of the periodic boundary conditions. 

Moreover, a Boltzmann approach was proposed by Bertin \etal \cite{Bertin2009}. Through binary collision integrals, a generalization of the Navier-Stokes equation was obtained and its consistency with the Toner and Tu equations was proved. A stability analysis of the stationary solutions showed that a stable homogeneous velocity field exists inside the ordered phase. The only instabilities found in the system were due to inhomogeneous perturbations, parallel to  a homogeneous non-zero velocity field. This point of view has the disadvantage that it only deals with binary collisions. Because a particle belonging to the MC has, on average, hundreds of neighbors, the binary collision approximation is expected to fail in describing the MC. 

For their part, Peruani \etal \cite{Peruani2008} used a Fokker-Planck generalization of a continuous version of Eq. \eqref{eq:langevin} to obtain the density function $\dens$ for the positions $\mathbf x$ and the orientations $\theta$ of the particles at time $t$. The resulting equation can be written as

\begin{equation}\label{eq:fokker}
 \dot{\psi} (\mathbf x,\theta,t) =  \partial_{\theta\theta} \dens - \partial_{\theta}[{F}_{\theta}] - \nabla [\mathbf{F_x}\dens]
\end{equation}

\noindent where the effects of diffusion, alignment and self-propulsion are described by the first, second and third terms in the r.h.s. respectively. Here $D_{\theta}$ is the rotational self-diffusion of the particles.  ${F}_{\theta}=-\gamma \int_{R(\mathbf{x})}d\mathbf{x}'\int_0^{2\pi} d\theta' \frac{\partial U(\mathbf{x},\theta,\mathbf{x}',\theta')}{\partial \theta} \psi(\mathbf{x}',\theta,t)$ describes the alignment force due to a potential $U(\mathbf{x},\theta,\mathbf{x}',\theta')$ that acts on pairs of agents in a radius $R(\mathbf{x})$  and $\mathbf{F}_x=v_0\mathbf{V}(\theta)$ is the self-propulsion force driving each agent.    

A perturbative treatment of the disordered state was proposed and the condition for the instability of the  homogeneous state,  $\psi_0 >\frac{2D_{\theta}}{\gamma\pi R_0^2}$, was found with $\gamma$ a damping constant. Hence, for a given noise intensity, expressed by $D_{\theta}$, there is a critical particle density above which the homogeneous solution is no longer stable.

There have been other attempts (Dedgond, Ramaswami, among others) to describe the ODT by density and velocity fields. A  review on this topic, authored by Bertin \cite{Bertin2016}, has recently appeared.

At first sight, these results seem to contradict the existence of just one cluster in a bounded region of space. This may occur because these models were implemented to \emph{ describe the order-disorder} transition in the Vicsek model. As a consequence, the stability of velocity fields that explains that particles move in the same direction, will not give any information about the onset of an MC formation.     

As it was proposed by Solon \etal \cite{Solon2015}, a stochastic approach could be most suitable to describe the whole picture because the phase separation in SPP is different from those  in equilibrium systems. In this line, a possibility could be to prove that starting with a density field in a compact region of  space, it becomes unstable when the noise intensity, density or speed is incremented, as it was done by Peruani \etal \cite{Peruani2008} to describe the instability of the homogeneous disordered phase. Theoretical work along this line is currently in progress.

\begin{figure}[h]
\centering
\includegraphics[width=0.4\textwidth]{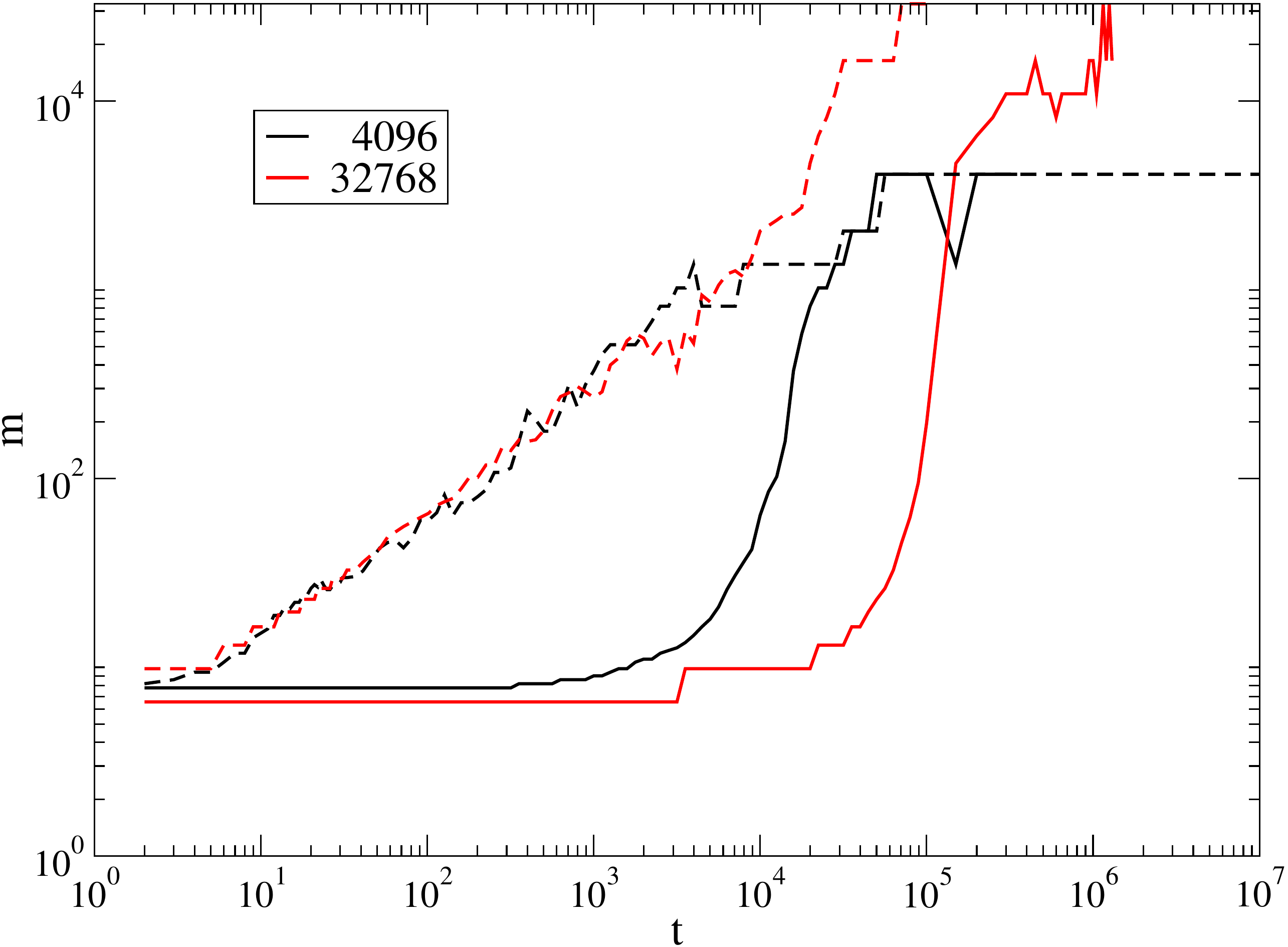}
 \caption{Cluster size evolution. The time to form an  MC from a specific initial condition  drastically changes with system size (continuous lines). This dependency vanishes when random initial conditions are set up (dashed lines).  }\label{fig:zero}
\end{figure}

Finally, a remark on the dependence on initial conditions: Since most simulations start  with a random distribution in space and orientation, larger system sizes will require longer physical times until the MC configuration emerges.  To elucidate this point an initial condition with particles randomly distributed in space but with the same orientation ($\phi=1$) was set up at zero noise. Then one randomly chosen particle is randomly twisted and the time to obtain the MC is measured. It is not a surprise that, as shown by the continuous lines in  Fig. \ref{fig:zero}, showing the unnormalized cluster size $m = Nm^*$ for one realization, a system of size $N=2^{15}$ takes a factor of ten longer than one of size $N=2^{12}$ to form the MC. On the other hand, a disordered initial condition (dashed lines) takes almost the same time to form the MC. This difference occurs because the twisted particle needs, in larger systems, to spread out the information on the new state that the system must reach to more particles in  a sort of chain reaction. If the initial condition is homogeneous, first there is a nucleation process where locally polarized clusters appear. Thus the chain reaction will occur over clusters instead of particles. Therefore, it takes more time to the  polarized initial condition  to form the MC, as shown from the first plateau of the continuous curves. In the disordered initial conditions all particles are sharing the information on their orientation to each other, thus the assemblage of the MC only depends on density and speed. Note that the zero noise state, with $\phi=1$, is an absorbing state; once it is reached it is not possible to escape from it. Consequently, if an arbitrary CSD is initially configured, there is no chance of obtaining an MC in absence of noise. 

A plausible mathematical explanation of this phenomenon appears in Solon \etal \cite{Solon2015}, where deterministic field equations were presented.  A disordered solution of these equations becomes unstable when density is lowered below some limit density $\rho_t$. This change corresponds to the apparition of the dotted line in the phase portrait in Fig. \ref{fig:r01-mmedio}. Interestingly, the solutions of  these deterministic  equations are strongly dependent on the initial conditions. However, when a noise term is added to them, the solutions to the modified equations become independent of the initial conditions. Hence noise plays a major role in describing the onset to an absorbing state as expected. 
An immediate consequence is that, in a description of the MC based on field equations, the initial condition dependency at zero noise shown in Fig. \ref{fig:zero} must appear. Therefore field equations must include a noise term that vanishes when noise intensity becomes zero in order to allow initial condition dependent solutions. In this scenario, the MC  is likely to be a stable solution of a branch bifurcation that depends on the initial conditions of the system.

\section{Summary}

The emergence of a single cluster that contains a macroscopic fraction of the particles in the system occurs in the hitherto unexplored regime of very low noise intensities it was presented through numerical simulations. 

The interplay between diffusion, self-propulsion and alignment forces is shown to regulate the relative strengths of the coarsening, coalescence and diffusion processes. 

An extension of the phase diagram for the Vicsek model was sketched to include the new phenomenology.

A phase transition characterized by large fluctuations of cluster related quantities was introduced through finite size-scaling analysis.

It was argued that a proper  description in terms of field equations must deal with the compact density field describing a macrocluster. Such a description should show that and MC must become unstable under an increase of the  density, speed or noise intensity leading to a homogeneous ordered phase. 

Note that the emergence of a single cluster is also of interest in biological systems where the number of individuals exhibiting collective behavior and its life span are both finite.   

\begin{acknowledgments}
 This work was supported by SECyT-UNC (Project 113/17) and CONICET (PIP 11220110100794), Argentina. The author is grateful to Drs. Fernando Peruani, Carlos Condat and Sergio Cannas for fruitful discussions. Simulations were carried out in the  CICADA cluster of the UNSA and the Licallo cluster of Observatoire Côte d' Azur, both at Nice. Part of this work was supported by a CONICET-Argentina postdoctoral grant. 
\end{acknowledgments}


\bibliographystyle{apsrev}

\end{document}